\documentclass[12pt]{article}
\include{epsf}

\newcommand{\eqn}[1]{(\ref{#1})}
\def\appendix#1{\addtocounter{section}{1}\setcounter{equation}{0}
\renewcommand{\thesection}{\Alph{section}}
\section*{
\thesection\protect\indent \parbox[t]{11.715cm} {#1}}
\addcontentsline{toc}{section}{Appendix\thesection\ \ \ #1} }
\newcommand{\real}{{\bb R}} 

\def\bra#1{\left\langle #1\right|}
\def\ket#1{\left| #1\right\rangle}
\def\hs#1#2{\left\langle #1\right|\left. #2\right\rangle}

\font\mybb=msbm10 at 12pt
\def\bb#1{\hbox{\mybb#1}}

\def\nn{\nonumber}

\newcommand{\Tr}[1]{\:{\rm Tr}\,#1}

\def\be{\begin{equation}}
\def\ee{\end{equation}}
\def\bea{\begin{eqnarray}}
\def\eea{\end{eqnarray}}

\newcommand{\del}{\partial}

\newcommand{\dm}[2]{\ket{#1}\bra{#2}}
\newcommand{\pt}{\varphi^{\rm Tay}}

\begin{document}
\begin{titlepage}
\begin{flushright}

\baselineskip=12pt
DSF--21--2003\\ hep--th/0306247\\
\hfill{ }\\
June 2003
\end{flushright}

\begin{center}

\baselineskip=24pt

{\Large\bf The Fuzzy Disc}

\baselineskip=14pt

\vspace{1cm}

{{\bf F.~Lizzi, P.~Vitale and A.~Zampini}}
\\[6mm]
 {\it Dipartimento di Scienze Fisiche, Universit\`{a} di
Napoli {\sl Federico II}\\ and {\it INFN, Sezione di Napoli}\\
Monte S.~Angelo, Via Cintia, 80126 Napoli, Italy}\\ {\tt
fedele.lizzi, patrizia.vitale, alessandro.zampini@na.infn.it}
\\[10mm]

\end{center}

\vskip 2 cm

\begin{abstract}
We introduce a finite dimensional matrix model approximation to
the algebra of functions on a disc based on noncommutative
geometry. The algebra is a subalgebra of the one characterizing
the noncommutative plane with a $*$~product and depends on two
parameters $N$ and $\theta$. It is composed of functions which
decay exponentially outside a disc. In the limit in which the size
of the matrices goes to infinity and the noncommutativity
parameter goes to zero the disc becomes sharper. We introduce a
Laplacian defined on the whole algebra and calculate its
eigenvalues. We also calculate the two--points correlation
function for a free massless theory (Green's function). In both
cases the agreement with the exact result on the disc is very good
already for relatively small matrices. This opens up the
possibility for the study of field theories on the disc with
nonperturbative methods. The model contains edge states, a fact
studied in a similar matrix model independently introduced by
Balachandran, Gupta and K\"urk\c{c}\"{u}o\v{g}lu.

\end{abstract}

\end{titlepage}

\section{Introduction}

Noncommutative Geometry~\cite{Connes,Landi,Madorebook,Ticos}
provides the possibility to describe also ordinary geometries
approximating the fields defined on them with finite (matrix)
algebras. The archetype of these \emph{fuzzy spaces} is the fuzzy
sphere~\cite{Madorefuzzysphere}. Along the same lines other fuzzy
spaces have been built \cite{GrosseStrohamer,Ramgoolam,thefuzz}.
There is also the fascinating possibility that the very structure
of spacetime at ultrashort length scales might be fuzzy, with a
dramatic drop of degrees of freedom at short distances with its
consequences on the ultraviolet structure. As additional example
matrix models in noncommutative geometry also provide a
way~\cite{AlbuquerquedeLyraTeotonio} to ask questions (for the
moment in a toy model) like the number of points or the average
dimension of the universe, questions which was unthinkable to ask
in the usual context.

The fuzzy approximation is in some sense similar to the usual
lattice one, in that a continuous action is substituted by a model
with a finite number of degrees of freedom, thus enabling the
possibilities to solve field theories with a approximate path
integral formalism. In the limit in which matrices become large
the approximation improves. However there are some fundamental
differences.  One advantage of the fuzzy approximation over the
usual lattice is that the basic symmetries of the space at hand
are preserved, thus fuzzy spheres are $SO(3)$ invariant, fuzzy
tori are translationally invariant etc. The original spaces are
recovered when the size of the matrices goes to infinity. In this
sense noncommutative geometry, and fuzzy spaces in particular,
naturally lead to matrix models~\cite{AMNS,BCMP}, which can
approximate both commutative and noncommutative spaces. The study
of field theory on these spaces with numerical methods has
started~\cite{AmbjornCatteral,Nishimnuraetc,dubliners} and seems
very promising. In particular the authors
of~\cite{LangmannSzaboZarembo} solve exactly, on the Moyal plane,
a matrix model which has similarities with the one presented in
this paper for the disc. The finite matrix model can also be
useful as a cutoff for renormalization, as advocated for example
in~\cite{Wulkenhaar}. The connections between finite matrix
algebras coming from a noncommutative plane and finite geometries
has been noted in~\cite{Polychronakos}  in the context of
noncommutative Chern--Simons theory~\cite{NairPolychronakos} as a
``quantum Hall droplet'', while the connection between the
projectors we use and the disc appears
in~\cite{PinzulStern,PinzulStern2}. A model which has several
points in common with ours has been independently developed by
Balachandran, Gupta and
K\"urk\c{c}\"{u}o\v{g}lu~\cite{BalGuptaKurkcuoglu} in a matrix
model which represents the edge currents of a Chern--Simons theory
on an infinite strip. The difference between our model and theirs
lies mainly in the way the limit is performed.

In this paper we present a fuzzy approximation to the
two--dimensional disc. We hasten to add that although the root of
the approximation lies in noncommutative geometry (or even
earlier, in quantization) the space of functions being
approximated is a normal \emph{commutative} space of functions. In
the limit in which matrices become infinite the noncommutativity
disappears\footnote{In case the field theory under consideration
is a nonabelian theory there can of course be other sources of
noncommutativity which will remain. It is also possible (depending
on the limit taken) to describe theories on noncommutative
spaces.}. Another important aspect of the model is that, while  in
the lattice approximation the functions, and hence the
correlations, can be calculated only at a finite set of points, in
the fuzzy approximation presented here approximate functions have
values at all points on the plane. We will argue that the matrix
model we present corresponds to a theory on a disc in a variety of
ways. We start in section~\ref{se:ncplane}  with a brief reminder
of the noncommutative plane in the form which is useful for our
purposes, which makes evident the map between functions on the
plane and infinite matrices. In section~\ref{se:subalgebra} we
introduce the fuzzy disc, a particular finite matrix subalgebra
and we argue that the corresponding functions have mainly support
on a disc and decrease exponentially outside it. As the size of
matrices increase the boundary becomes sharper. We also discuss
the form that the ultraviolet--infrared mixing takes in this case
and argue the presence of edge states. In order to show that the
algebra we are proposing actually corresponds to a disc, in
section~\ref{se:Laplacian} we introduce derivatives and Laplacians
and compare the spectrum of the latter in the continuum and in the
finite approximation. In section~\ref{se:field} we sketch an
approach to field theories on a disc based on the fuzzy
approximation. Although the construction is well suited for
Montecarlo techniques, the correlation of a free massless scalar
field can be solved analytically giving the Green's function in
terms of the inverse of a matrix. We calculate it numerically and
compare it with the exact case. The final section contains
conclusions and outlook. Some technical details of calculations
are in the appendix.

\section{The Noncommutative Plane as a Matrix Algebra \label{se:ncplane}}
\setcounter{equation}{0}

We start from functions on $\real^2$, with coordinates $x$ and
$y$.  Consider this plane to be like a `quantum phase space', that
is that we ``quantize'' $x$ and $y$ and give them a nontrivial
commutator:
\be
[\hat x,\hat y]=i\frac\theta 2 \ .
\ee
It is convenient to consider the plane as a complex space with
$z=x+iy$. The quantized versions of $z$ and $\bar z$ are the usual
annihilation and creation operators, $a=\hat x +i \hat y$ and
$a^\dagger=\hat x -i \hat y$ with a slightly unusual
normalization, so that their commutation rule is
\be
[a,a^\dagger]=\theta \ . \label{aadcomm}
\ee
The parameter $\theta$, which has the dimensions of a square
length, plays the role of $\hbar$, but it has no physical meaning,
much like the distance between sites in a lattice approximation.

This ``quantization'' associates operators to functions. There are however
ambiguities in this association and in order to give a unambiguous
procedure we introduce a map which associates an operator on an Hilbert
space to a function on the plane. This is a quantization procedure and it
was essentially introduced by Weyl long ago~\cite{Weyl}. Given a function
on the plane we define a map $\Omega_\theta$ which to the function
$\varphi(z,\bar z)$ associates the operator $\hat \varphi$ defined as
follows. Given the function $\varphi(\bar z,z)$ consider its Taylor
expansion:
\be
\varphi(\bar z,z)=\pt_{mn}\bar z^m z^n  \ , \label{taylorphi}
\ee
to this function we associate the operator
\be
\Omega_\theta(\varphi):=\hat\varphi=\pt_{mn}{a^\dagger}^m a^n  \ ,
\label{taylorexp}
\ee
thus we have ``quantized'' the plane using a normal ordering
prescription. To be precise this map is not the one introduced by
Weyl (which associates hermitian operators to real functions) but
can be cast in the same language since
\be
\Omega_\theta(\varphi)=\int \frac{d^2u}{2\pi\theta} \hat
\varphi(\bar u, u)\, e^{ua^{\dagger}/\theta}e^{-\bar u a/\theta} \
,
\ee
where $\hat \varphi(\bar u, u)$ is the Fourier transform of
$\varphi$.

Because of the similarity of its integral expression to the
standard Weyl map we will also refer to it as the weighted Weyl
map. The map $\Omega_\theta$ is invertible. Its inverse is a
variation of the Wigner map~\cite{Wigner} (the inverse of the Weyl
map). It can be efficiently expressed defining the \emph{coherent}
states:
\be
a\ket{z}=z\ket{z}  \ ,
\ee
then it results
\be
\Omega^{-1}_\theta(\hat\varphi)=\varphi(\bar z,z)=\bra{z}\hat
\varphi\ket{z}  \ . \label{Omegam1}
\ee
There is another useful basis on which it is possible to represent
the operators. Consider the number operator
\be
{\rm N}=a^\dagger a  \ , \label{defnumb}
\ee
and its eigenvectors which we indicate\footnote{There is a
possible notational confusion here. While it is true that the
coherent state for $z=0$ is the same as the eigenvector of ${\rm
N}$ with zero eigenvalues, it is not true that the coherent state
$\ket{z}|_{z=n}$ is the same as $\ket{n}$. Since we never consider
coherent states with integer values we refrain from the
introduction of another symbol for the eigenvectors of ${\rm N}$.}
by $\ket{n}$:
\be
{\rm N}\ket{n}=n\theta\ket{n}  \ .
\ee
We can then express the operators within a density matrix notation:
\be
\hat\varphi=\sum_{m,n=0}^\infty\varphi_{mn}\dm{m}{n}  \ .
\label{dmexp}
\ee
The elements of the density matrix basis have a very simple multiplication
rule:
\be
\ket{m}\hs{n}{p}\bra{q}=\delta_{np}\dm{m}{q}  \ . \label{densmult}
\ee
The connection between the expansions~\eqn{taylorexp}
and~\eqn{dmexp} is given by:
\bea
a&=&\sum_{n=0}^\infty \sqrt{(n+1)\,\theta}\dm{n}{n+1}\nn\\
a^\dagger&=&\sum_{n=0}^\infty \sqrt{(n+1)\,\theta}\dm{n+1}{n} \ .
\eea
Some useful conversion formulas are collected in
appendix~\ref{appcon}. Note that with our conventions $x,y,z,a$
and $a^\dagger$ have the dimension of a length,  ${\rm N}$ and
$\theta$ those of a length square, while density matrices are
dimensionless. Applying the dequantization map \eqn{Omegam1} to
the operator $\hat\varphi$ in the number basis we obtain for the
function $\varphi$ a new expression, analogous to the Taylor
expansion \eqn{taylorphi} in terms of the coefficient
$\varphi_{nm}$, that is
\be
\varphi(\bar
z,z)=e^{-\frac{|z|^2}{\theta}}\sum_{m,n=0}^\infty\varphi_{mn}
\frac{\bar z^m z^n}{\sqrt{n!m!\theta^{m+n}}}  \ , \label{newtay}
\ee
where we used the expression~\eqn{znsca}. The ordinary Wigner map
of these density matrix elements (the Wigner functions) can be
found in~\cite{Zachos}.

The maps $\Omega$ and $\Omega^{-1}$ yield a procedure of going
back and forth from functions to operators. Moreover,  the product
of operators being noncommutative,  a noncommutative $*$ product
between functions is implicitly defined as
\be
\left(\varphi*\varphi'\right)(\bar z,
z)=\Omega^{-1}\left(\Omega(\varphi)\, \Omega(\varphi')\right)  \ .
\label{starint}
\ee
This product (which is a variation of the Moyal-Gr\"{o}newold
product~\cite{Moyal} at the basis of   deformation quantization) was first
introduced by Voros \cite{Voros}. Its differential expression is
\be
\left(\varphi*\varphi'\right)(\bar z, z)=\left.e^{\theta\del_{\bar
z'}\del_{
z"}}\varphi(z')\varphi(z")\right|_{z=z'=z"}=\varphi\varphi'+\theta\del_{\bar
z}\varphi'\del_{z}\varphi'+O(\theta^2)  \ . \label{stardiff}
\ee
We will indicate the algebra of functions on the plane with this
product as ${\cal A}_\theta$.

In the density matrix basis, because of~\eqn{densmult}, the
product~\eqn{stardiff} simplifies to an infinite row by column matrix
multiplication:
\be
\left(\varphi*\varphi'\right)_{mn}=\sum_{k=1}^\infty
\varphi_{mk}\varphi'_{kn} \ .
\ee
Although  the map $\Omega$ is not defined for $\theta=0$, using
equation~\eqn{stardiff}, it is easy to see that, when
$\theta\to0$,  the $*$ product goes to the ordinary commutative
product. A word of caution, the various maps and products defined
here have domains and ranges which are not identical. While the
standard Weyl map maps Schwarzian functions into Hilbert Schmidt
operators, for the weighted Weyl map~\eqn{taylorexp} this is not
always the case.

Using relation~\eqn{znsca} or directly \eqn{newtay} it is easy to
see that also:
\be
\int d^2z\, \varphi(\bar z,z)={\pi\theta}
\Tr\Phi={\pi\theta}\sum_{n=0}^\infty\varphi_{nn} \ ,
\label{defint}
\ee
where we have introduced the matrix $\Phi$ with components
$\varphi_{nm}$.

\section{The Fuzzy Disc Subalgebra \label{se:subalgebra}}
\setcounter{equation}{0}

The main point of the previous section is that the map $\Omega$
and its inverse provide a manner to associate to each function,
$\varphi$  an infinite dimensional matrix $\Phi$. The price to pay
is that the commutative product of functions gets deformed through
a  parameter $\theta$. In this section we define the subalgebras
(with respect to the Voros $*$ product), of \emph{finite}
$N{\times} N$ matrices. Considering the functions whose
expansion~\eqn{newtay} terminates when either $n$ or $m$ is larger
than a given integer $N$, it is immediate to see that these
functions are closed under $*$ multiplication. In the limit of
$N\to\infty$, with $\theta$ fixed, the noncommutative plane is
recovered.

\begin{figure}[htbp]
\epsfxsize=2.5 in
\bigskip
\centerline{\epsffile{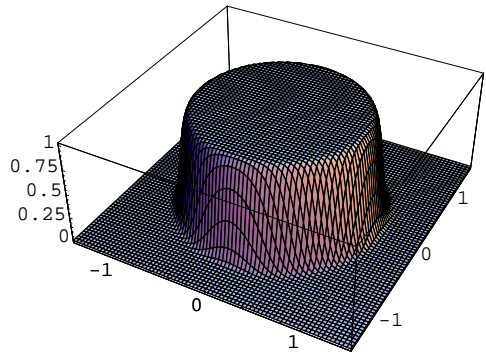}} \caption{\baselineskip=12pt
{\it The function $P^N_\theta$ for $N=10^2$.}}
\bigskip
\label{disc3d}
\end{figure}

The subalgebra can be obtained easily from the full algebra of
functions via a projection using the idempotent function
introduced in a similar context by Pinzul and
Stern~\cite{PinzulStern2}:
\be
P^N_\theta=\sum_{n=0}^N \hs{z}{n}\hs{n}{z}=\sum_{n=0}^N\frac{
r^{2n}}{n!\theta^n}e^{-\frac{r^2}{\theta}}  \ ,
\ee
where we have used the polar decomposition of $z=r e^{i\phi}$
and~\eqn{znsca}. The finite sum may be performed thus yielding
\be
P^N_\theta=\frac{\Gamma(N+1, {r^2\over \theta})}{\Gamma(N+1)}  \ ,
\ee
where $\Gamma(n,x)$ denotes  the  incomplete gamma function. In
this notation it is then clear that, in the limit $N\to \infty$
and $\theta\to 0$ with
\be
R^2\equiv N\theta
\ee
fixed, the sum converges to $1$ if $r^2/\theta<N$ (namely $r<R$),
and converges to 0 otherwise. It has cylindrical symmetry since
$\phi$ does not appear. For $N$ finite the function vanishes
exponentially for $r$ larger than $R$ (see figure~\ref{disc3d}).

\begin{figure}[htbp]
\epsfxsize=2.5 in
\bigskip
\centerline{\epsffile{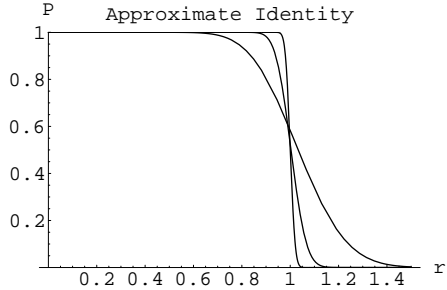}} \caption{\baselineskip=12pt
{\it Profile of the spherically symmetric function $P^N_\theta$ for the
choice $R^2=N\theta=1$ and $N=10, 10^2, 10^3$. As $N$ increases the step
becomes sharper.}}
\bigskip
\label{identity}
\end{figure}

Already for $N= 10^3$ it is well approximated (see
figure~\ref{identity}) by a step function.

The function $P^N_\theta$ is a projector of the algebra of
functions on the plane with the $*$ product:
\be
P^N_\theta*P^N_\theta=P^N_\theta  \ ,
\ee
and the subalgebra ${\cal A}^N_\theta$ is defined as
\be
{\cal A}_\theta^N=P^N_\theta*{\cal A}_\theta*P^N_\theta  \ .
\ee
Cutting at a finite $N$ the expansion provides an infrared cutoff.
The cutoff is fuzzy in the sense that functions in the subalgebra
are still defined outside the cutoff, but are exponentially
damped. For example a Gaussian of width $\theta$ (corresponding to
the element $\dm{0}{0}$) when projected, is mapped into itself,
and is nonzero on the whole plane. The map for a normalized
Gaussian centred around the origin but with width $\alpha>\theta$
also shows the infrared cutoff. Let us investigate the effect of
$P^N_\theta$ on some functions. We use the notation
\be
\Pi^N_\theta(\varphi)\equiv
P^N_\theta*\varphi*P^N_\theta=\sum_{m,n=1}^\infty
\varphi_{mn}e^{-\frac{|z|^2}{\theta}}\frac{\bar z^m
z^n}{\sqrt{m!n!\theta^{m+n}}}
\ee
and talk of ``fuzzy functions'' for the projected functions so
that $P^N_\theta=\Pi^N_\theta(1)$ is the fuzzy identity. At the
level of operators $P^N_\theta$ corresponds to the projector
\be
\hat P^N_\theta=\sum_{n=0}^N \dm{n}{n}  \ .
\ee\
We call \emph{fuzzy disc} the space corresponding to the algebra
${\cal A}_\theta^N$, which is isomorphic to the algebra of
$N\times N$ matrices. This, as a matrix algebra, is the same as
the one for the fuzzy torus, the fuzzy disc and any fuzzy spaces
in general. In~\cite{HLS-J} the fuzzy sphere is indeed obtained
with the use a projector on the Voros algebra of $\real^3$. What
makes it a disc are the way to take the correlated limit of
$\theta$ and $N$ keeping the dimensionful quantity $R$ fixed,
which we now pass to describe, and the extra structures we will
discuss in section~\ref{se:Laplacian}.

The limit discussed here must be understood in an heuristic sense,
since the algebra of functions on the disc is not approximatively
finite one would have to consider it in a weak sense, like the way
it is done in~\cite{LandiLizziSzaboLargeN}. A more rigorous
analysis on the line of what is done in~\cite{Rieffel} would also
be desirable. It must also be noted that different ways to take
the limit give different commutative or noncommutative spaces, as
discussed for example in~\cite{VadiyaYdri,PinzulStern}.

We first calculate the effect of the projector on a rotationally
symmetric Gaussian centred at the origin of width $\alpha$. Any
cylindrically symmetric function $\varphi(r)$ has a simple
expansion:
\be
\varphi(r)=e^{-\frac{r^2}\theta}\sum_{n=0}^\infty
\varphi_n\frac{r^{2n}}{\theta^n n!}  \ ,
\ee
where $\varphi_n$ can be calculated from~\eqn{taylortodm}. In
particular for the normalized Gaussian
\be
\varphi(r)=\frac{1}{\pi\alpha}e^{-\frac{r^2}{\alpha}}
\label{ngaus}
\ee
we have
\be
\varphi_n=\frac{1}{\pi\alpha}\left(1-\frac\theta\alpha\right)^n  \
, \label{gaussseries}
\ee
and the series can be summed to give
\be
\Pi^N_\theta(\varphi(r))=e^{-\frac{r^2}\theta}\sum_{n=0}^N\varphi_n
\frac{r^{2n}}{n!\theta^n}=e^{-\frac{r^2}\alpha}\frac{\Gamma\left(N+1,r^2\left(\frac
1\theta-\frac 1\alpha\right)\right)}{\pi\alpha \Gamma(N+1)} \ .
\label{gaussorigin}
\ee
The map $\Pi^N_\theta$ is just an infrared cutoff which in the
limit becomes sharper and sharper.
\begin{figure}[htbp]
\epsfxsize=2.3 in \centerline{\epsfxsize=2.2
in\epsffile{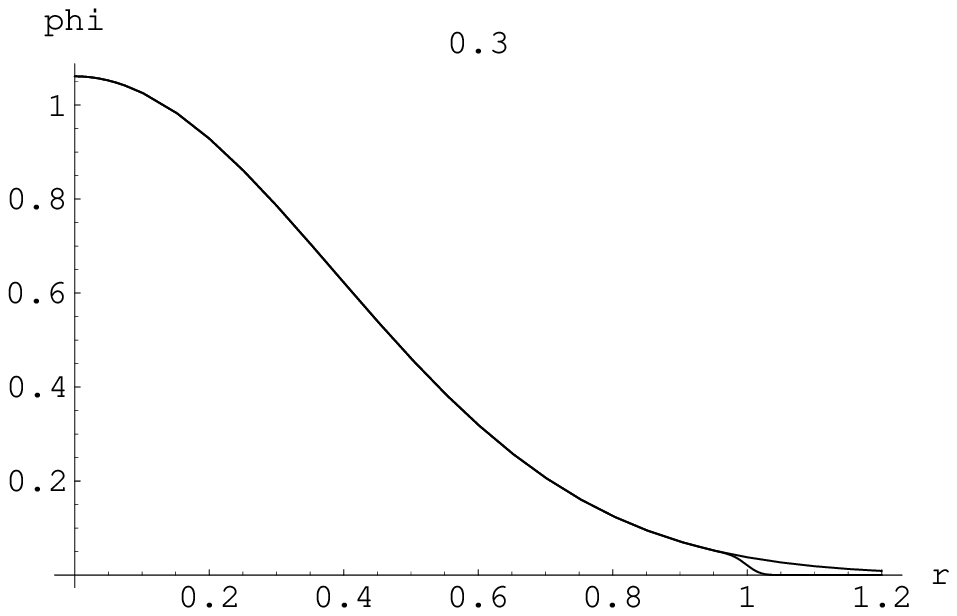}\epsfxsize=2.2
in\epsffile{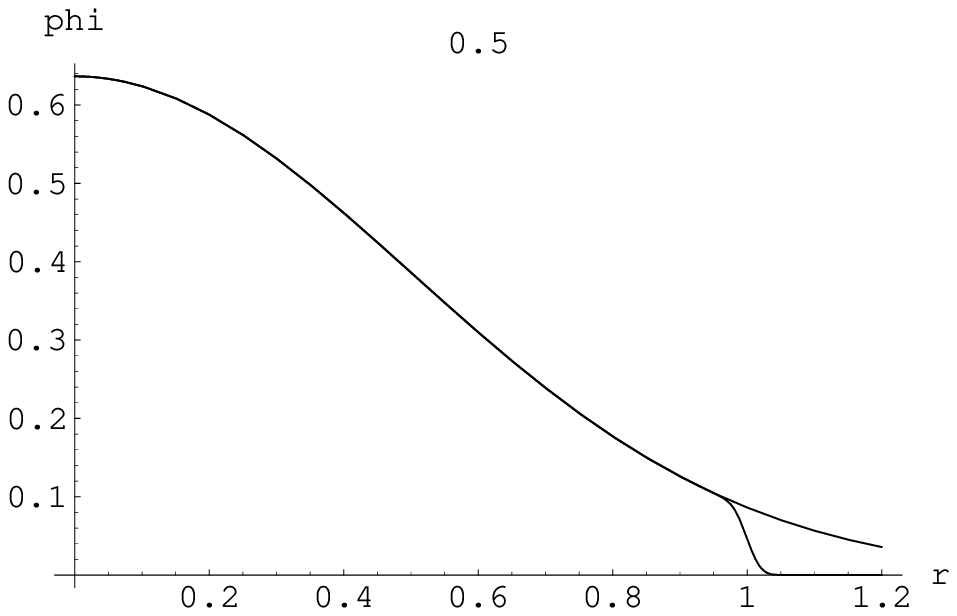}\epsfxsize=2.2 in\epsffile{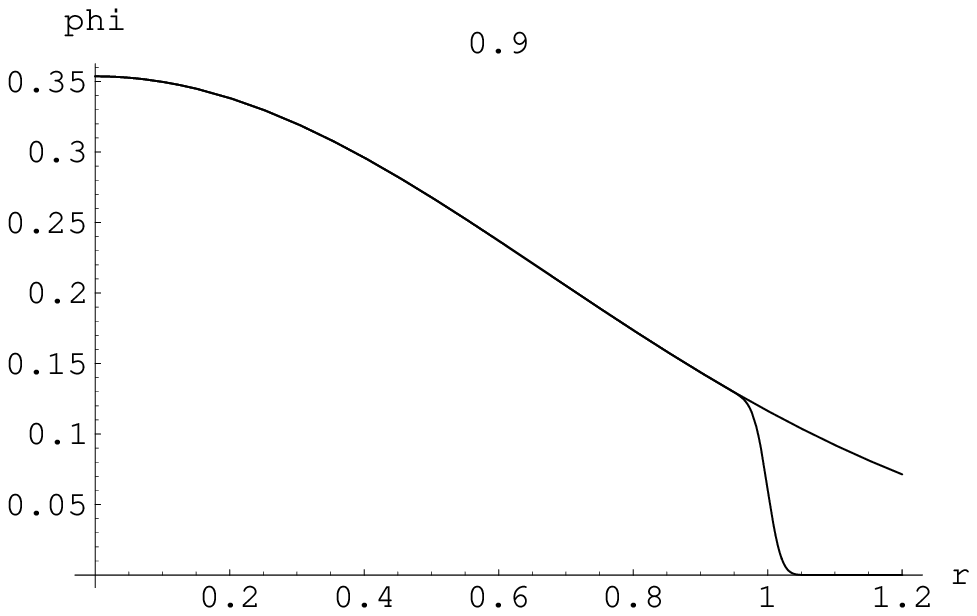}}
\caption{\baselineskip=12pt {\it Profile of the spherically
symmetric function
$\Pi^N_\theta\left(\frac{1}{\pi\alpha}e^{-\frac{r^2}{\alpha}}\right)$
for the choice $R^2=N\theta=1$, $N=10^3$ and for various choices
of $\alpha$ compared with the unprojected function. Both functions
are plotted, although inside the disc they are practically
indistinguishable. The unprojected function is always the larger
one.}}
\bigskip
\label{figgaussorigin}
\end{figure}
Equation~\eqn{gaussorigin} is plotted in
figure~\ref{figgaussorigin} for $\theta<\alpha$.

\begin{figure}[tbph]
\epsfxsize=2.5 in \centerline{\epsfxsize=2.2
in\epsffile{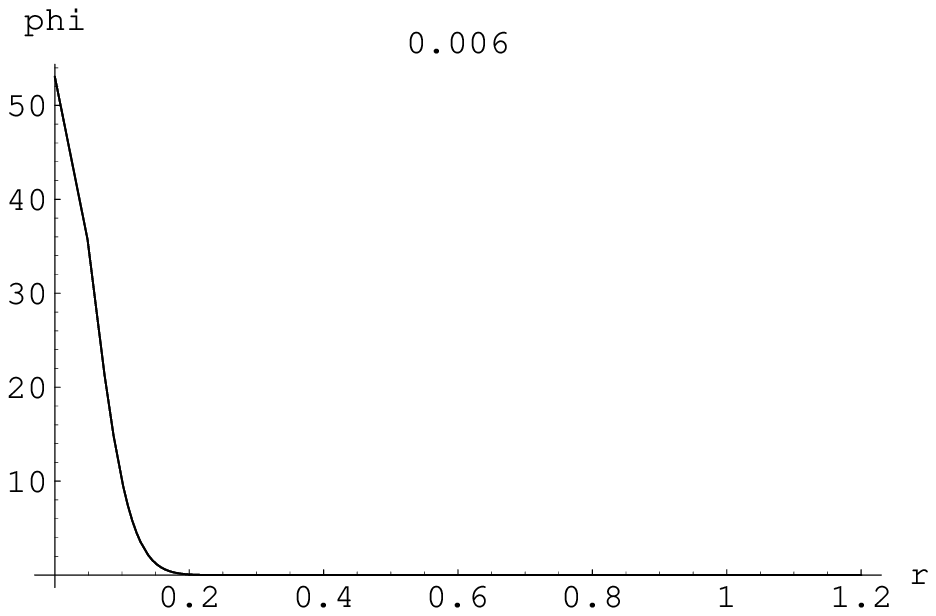} \epsfxsize=2.2
in\epsffile{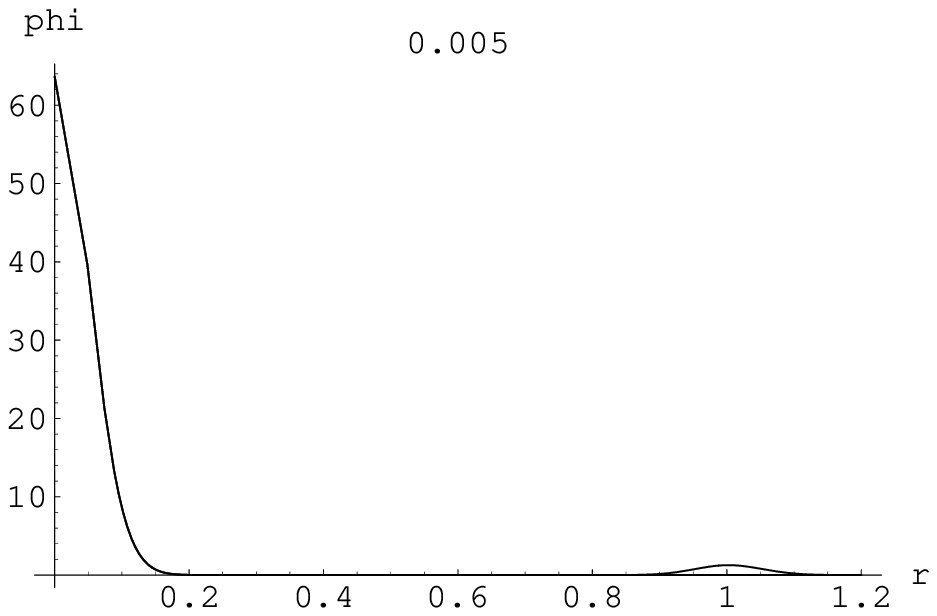}\epsfxsize=2.2
in\epsffile{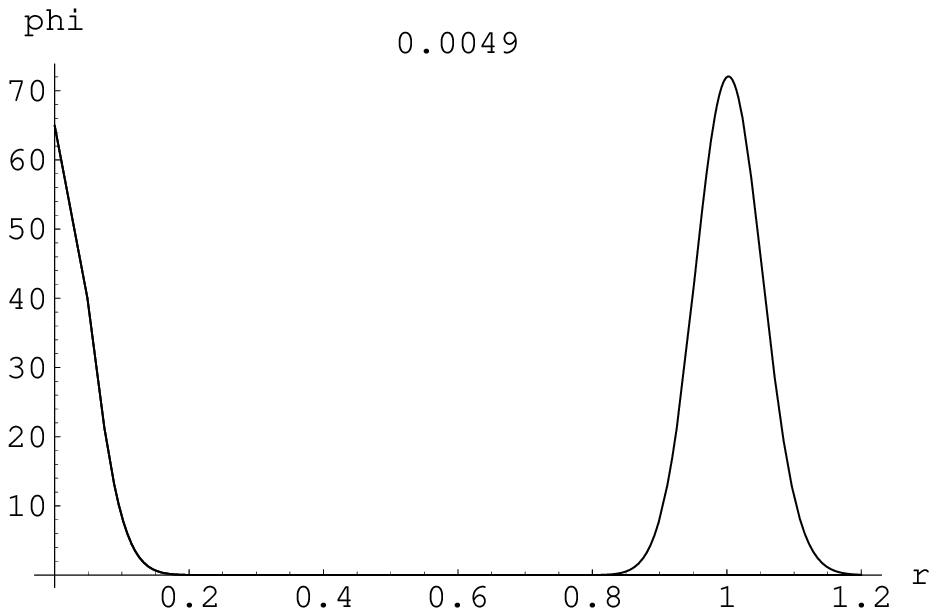} } \caption{\baselineskip=12pt {\it
Profile of the spherically symmetric function
$\Pi^N_\theta(\left(\frac{1}{\pi\alpha}e^{-\frac{r^2}{\alpha}}\right))$
for the choice $R^2=N\theta=1$, $N=10^2$ and for various choices
of $\alpha$.}}
\bigskip
\label{figorigingauss}
\end{figure}

Things are very different if we try to localize the function at a
distance smaller than $\sqrt{\theta}$. Figure~\ref{figorigingauss}
shows what happens in such case: for the larger value, $\alpha=.6
\theta$, the projected function is undistinguishable from the
exact function (both are actually plotted in the figure). For the
middle value, $\alpha=.5 \theta$, with our numerical
approximations, a small ``bump'' at $r=R$ appears. Already for
$\alpha=.49 \theta$ the 'bump' has become a large Gaussian sitting
at the infrared cutoff; the part close to the origin is still
there, but it is quickly dwarfed by the infrared bump which grows
very fast. For $\alpha\sim .4$ it is already of the order of
$10^{17}$. Keeping $\alpha$ fixed and increasing $N$, with $R$
fixed, the bump disappears. The reason for this behaviour is that
the factors $\varphi_n$ in~\eqn{gaussseries} become negative and
smaller than~-1. Therefore the individual terms of the series
become larger and larger. Keeping the whole series provides
cancellations which do not happen truncating the series at a
finite value. In fact the bump becomes negative for $N$ odd. This
is a reflection of the fact that, with the weighted Weyl map, the
operator corresponding to the Schwarzian function~\eqn{ngaus} with
$\alpha\leq\theta/2$ is not compact, and hence it cannot be
approximated by finite rank matrices.

In general the function $\varphi$ is close to
$\Pi^N_\theta(\varphi)$ if it is mostly supported on the disc of
radius $R$ (otherwise it is simply exponentially cut) and if it
does not have oscillations of wavelength smaller than
$\sqrt{\theta}$. It this case the function becomes very large on
the boundary. This is a \emph{compact} example of the
\emph{ultraviolet-infrared}
mixing~\cite{MinwallaSeiberVanRamsdoong} which is one of the most
interesting characteristics of noncommutative geometry. If we try
to localize too much the function, unavoidably an infrared
divergence on the boundary of the disc appears.

\begin{figure}[htbp]
\epsfxsize=2.5 in \centerline{\epsfxsize=2.5
in\epsffile{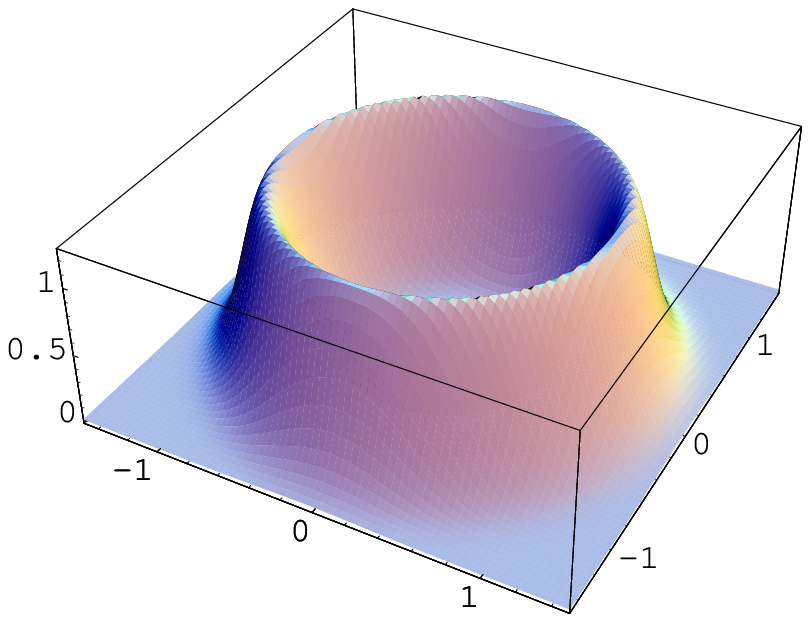} \epsfxsize=2.5
in\epsffile{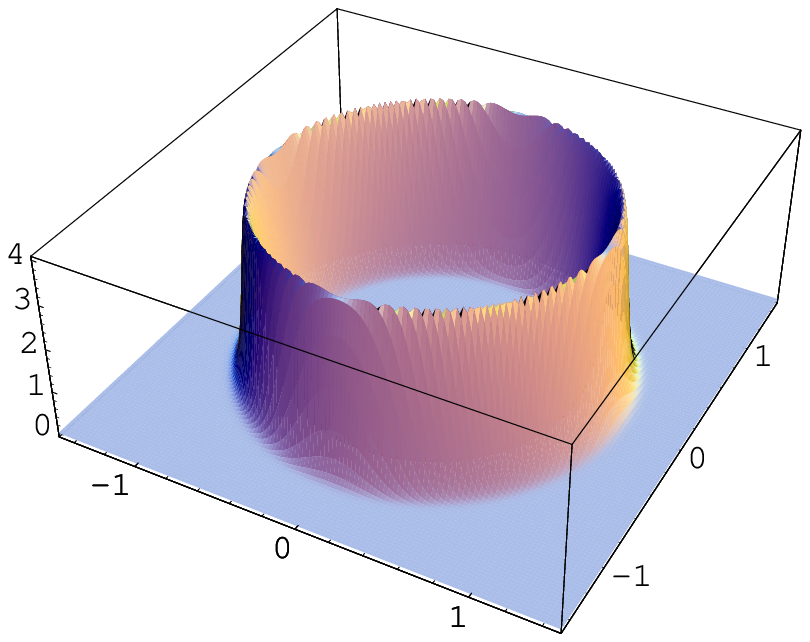}} \caption{\baselineskip=12pt {\it
The edge states $\varphi^{\mbox{\rm edge}}$ for $N=10$ and
$N=100$.  }}
\bigskip
\label{edge}
\end{figure}

There are however functions which are localized sharply near the
edge of the disc. These are the edge states~\cite{edge} which play
an important role in Chern-Simons theory and have been introduced
in these matrix models  by A.P.~Balachandran, K,~Gupta and
S.~K\"urk\c{c}\"{u}o\v{g}lu in~\cite{BalGuptaKurkcuoglu}, and the
discussion presented here is inspired by this work. The edge
states are simply given by the highest one--dimensional projector:
\be
\varphi^{\mbox{edge}}=\frac{1}{\theta}\hs{z}{N}\hs{N}{z}=e^{-\frac{r^2}\theta}
\frac{r^{2N}}{\theta^{N+1} N!}  \ .
\ee
They are plotted in figure~\ref{edge} for $N=10$ and $N=100$. It
is evident that as $N$ increases they become sharper.

\section{Fuzzy Derivatives and Fuzzy Laplacians \label{se:Laplacian}}
\setcounter{equation}{0} So far we have defined a projection from
functions on the plane to a finite dimensional algebra of $N
{\times} N$ matrices and discovered that, with an appropriate
choice of $N$ and $\theta$ we obtain a good approximation of
functions supported on a disc. But, in the spirit of
noncommutative geometry, if we want to talk of geometry, we have
to define a Dirac operator or a Laplacian which give the metric
properties of the system~\cite{Connes}. To define derivatives thus
we define a pair of operators which act on the same space as the
elements of ${\cal A}_\theta^N$. Their commutators with a function
define what we usually call the derivatives of a function.  Of
course we require that in the limit these two derivations converge
to the usual derivatives of the disc. Combining them with $\gamma$
matrices would give the Dirac operator, to avoid discussing
fermions we will concentrate on the Laplacian.

The starting point to define the matrix equivalent of the
derivations is:
\bea
\del_z\varphi&=&\frac{1}{\theta}\bra{z}
[a^\dagger,\Omega(\varphi)]\ket{z}\nn\\
\del_{\bar
z}\varphi&=&\frac{1}{\theta}\bra{z}[a,\Omega(\varphi)]\ket{z}  \ .
\eea
The above expression is exact. Acting on an element of the
subalgebra ${\cal A}_\theta^N$ the derivative takes the functions
out of the algebra, a phenomenon not uncommon in noncommutative
geometry. However
\be
\del_z({\cal A}_\theta^N)\subset {\cal A}_\theta^{N+1}
\ee
analogously for $\del_{\bar z}$, so that the derivatives of
functions of the subalgebra can still be considered finite
matrices. Notice that
\be
\del_z\Pi^N_\theta(\varphi)\neq \Pi^N_\theta(\del_z\varphi)  \ ,
\ee
the equality obviously holding in the limit. We will use
$\del_z\Pi^N_\theta(\varphi)$ to define the Laplacian below.

The fact that we keep $a$ and $a^\dagger$ operators on the full
space (hence they are still infinite matrices) is crucial for the
identification of the algebra of $N\times N$ matrices with the
approximation to the disc. If we were just to truncate the
matrices $a$ and $a^\dagger$, their commutator would not be
proportional to the identity. For the fuzzy torus the derivations
cannot be expressed as commutators, while for the fuzzy sphere
there are three derivations which are not however independent, but
are connected by a Casimir. This goes just to say that the same
algebra, with different structures, can approximate different
spaces.

Rotations are well defined, in fact the generator of rotations on
the fuzzy disc is nothing but the number operator ${\rm N}$
introduced in~\eqn{defnumb} which commutes with $\hat P_\theta^N$
just as the generator of rotations on the ordinary disc, the
angular momentum operator $\del_\phi$, commutes with $P_\theta^N$.
Just as the fuzzy sphere maintains the invariance group of the
sphere, the fuzzy disc retains the fundamental symmetry of the
disc.

Note that the eigenvalue equations
\bea
\frac 1\theta [a,\varphi]=\lambda \varphi\nonumber\\
\frac 1\theta [a^\dagger,\varphi]=\lambda \varphi
\eea
have no solution in the space of $N\times N$ matrices, just as in
the commutative case translations on the disc have no
eigenvectors. Nevertheless the fuzzy Laplacian
\be
\hat\nabla^2\hat\varphi:=\frac{4}{\theta^2}[a,[a^\dagger,\varphi]]
=\frac{4}{\theta^2}[a^\dagger,[a,\varphi]] \label{lapladef}
\ee
can be defined. In particular consider the following matrix model:
\be
S=\frac 1{2\pi\theta}\int d^2z\; \varphi^* *(\nabla^2\varphi)
=\Tr\Phi_{mn}^\dagger \left(\nabla^2\right)_{mnpq}\Phi_{pq}  \ ,
\ee
where $\left(\nabla^2\right)_{mnpq}$ is implictly defined
by~\eqn{lapladef}. It is an operator acting on a finite space of
dimension $(N+1)^2$, and its eigenvalues can be calculated and
compared with the exact commutative case.

\begin{figure}[htbp]
\epsfxsize=2.5 in \centerline{\epsfxsize=2.5
in\epsffile{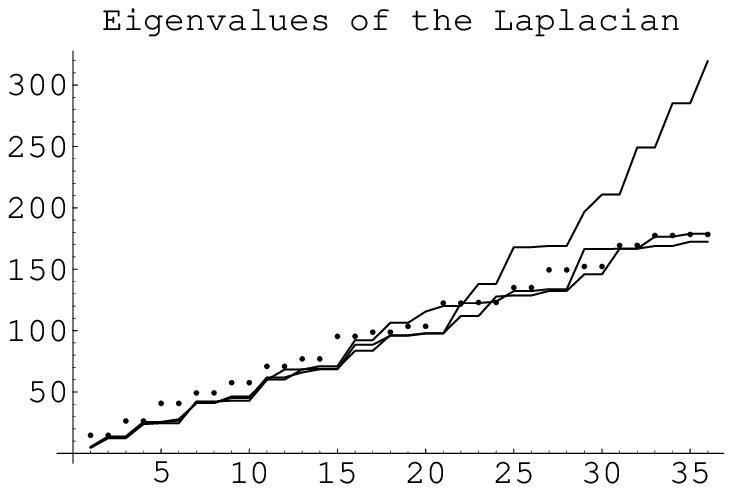} \epsfxsize=2.5 in\epsffile{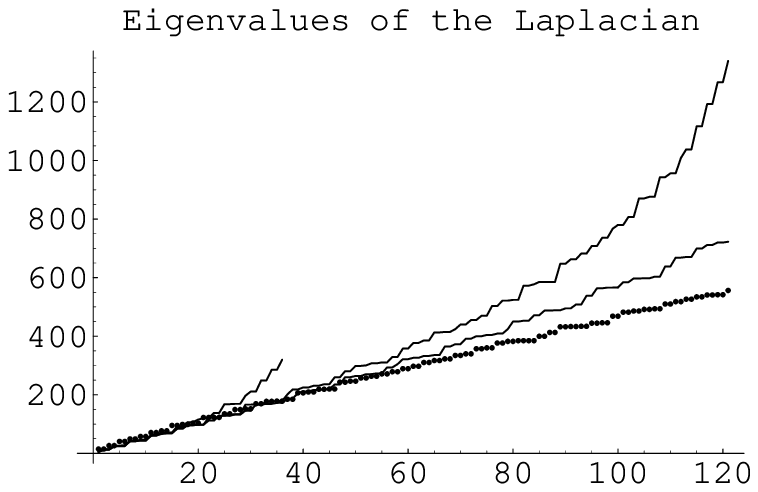}}
\caption{\baselineskip=12pt {\it The first eigenvalues of the
Laplacian on the disc (dots) and the fuzzy Laplacian (solid lines) for
$N=5, 10,15$. The lines corresponding to the three cases can be
distinguished because the agreement with the exact case improves as $N$
grows. In the figure on the right the curve which interrupts is the one
corresponding to $N=5$, for which there are only 36 eigenvalues. }}
\bigskip
\label{figlapl}
\end{figure}

As is well known the eigenfunctions of the Laplacian on a disc
with Dirichlet boundary conditions are the Bessel and the Neumann
functions of integer order multiplied by $\exp (im\phi)$. The
eigenvalues, $E_{mn}$, are related to the zeroes of these
functions. More precisely, we have
\be
E_{mn}= ({zeroes \over R})^2  \ .
\ee
In figure~\ref{figlapl} we show a comparison between the
eigenvalues for the exact and approximate Laplacians for three
values of $N$.
The agreement is fairly good, a fuzzy drum sounds pretty much like
a regular drum. Discrepancies start occurring  for the $4 N^{th}$
eigenvalues, this is to be expected because $4 N / R^2=4/\theta$
is the energy cutoff of the theory\footnote{We use units for which
$\hbar=c=1$}. Most of the eigenvalues (but not all) of the fuzzy
Laplacian are doubly degenerate, but the unmatched ones become
sparser as $N$ increases.
\section{Free Field Theory on the Fuzzy Disc: Green's functions
\label{se:field}}
\setcounter{equation}{0}
The formalism we have developed lends itself readily for matrix
approximations to field theories on a disc. For the real scalar
case described by the action
\be
S=\int d^2z\; \varphi \nabla^2 \varphi + \frac{m^2}{2}\varphi^2
+V(\varphi)  \ ,
\ee
we consider the fuzzy action
\be
S_\theta^2=\frac{1}{\pi} \Tr \Phi \hat\nabla^2 \Phi +
\frac{m^2}{2} \Phi^2 +V(\Phi)  \ .
\ee
We stress that this is an action entirely of finite dimensional
matrices, which can be approached numerically, using for example
Montecarlo techniques, a method currently in use for other fuzzy
spaces~\cite{AmbjornCatteral,Nishimnuraetc,dubliners}. Here we
will content ourselves in calculating the two points Green
function for the free massless scalar theory. In this case the
path integral may be performed yielding just the inverse of the
Laplacian with Dirichlet boundary conditions which we have
calculated in section~\ref{se:Laplacian}:
\be
\left\langle\varphi (\bar z, z)\varphi (\bar z', z')\right\rangle=
G(z,z')=\bra{z}(\nabla^2)^{-1}\ket{z'}  \ .
\ee
In the continuous theory on the disc this can be calculated
exactly using standard techniques of classical electrodynamics(see
for example~\cite{smirnov}) yielding
\be
G(z,z')= {1 \over 2 \pi} \ln \frac{|z-z'|}{|z'|z|-z|z|^{-1}|} \ .
\label{exactgre}
\ee
The fuzzy approximation is
\be
G^{(N)}_\theta (z,z')= 4 \sum^N_{m,n,p,q =1}
\frac{e^{-{|z|^2+|z'|^2\over \theta}} (\nabla^{-2})_{mnpq}\bar z^p
z^q z'^m \bar z'^n }{\sqrt{p!q!m!n!\theta^{m+n+p+q}}} \ .
\label{fuzzygre}
\ee
This expression can be evaluated numerically.
\begin{figure}[thb]
\epsfxsize=2.5 in \centerline{\epsfxsize=2.5
in\epsffile{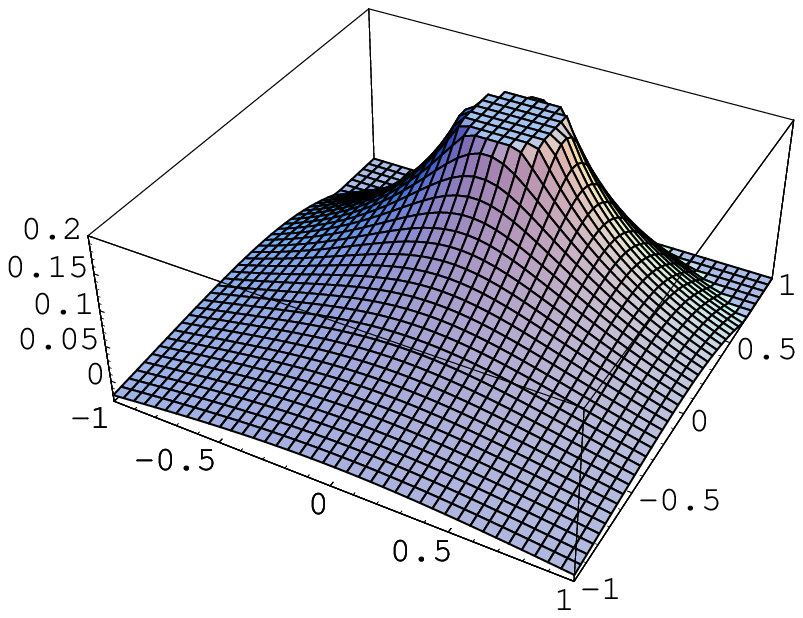} \epsfxsize=2.5 in\epsffile{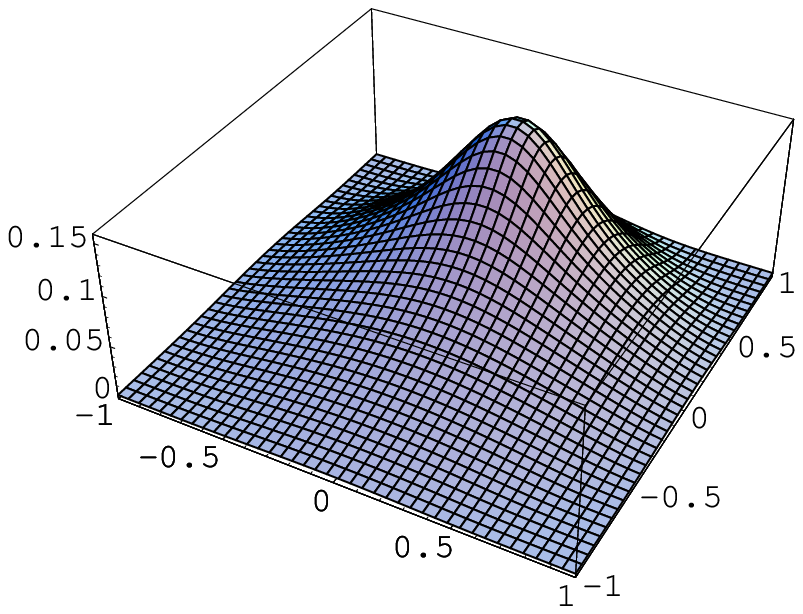}}
\caption{\baselineskip=12pt {\it Comparison between the Green's
function $G(z,z')$ as function of $z$ for $z'=i/2$ On the left the
exact function (with the singularity at $z=z'$ truncated), on the
right the approximated function for $N=20$ . }}
\bigskip
\label{3d}
\end{figure}
In figures~\ref{3d} and~\ref{densityplot} we show a comparison
between the exact Green function on the disc~\eqn{exactgre}
and~\eqn{fuzzygre}.

The agreement is, in our opinion, quite remarkable already for a
limited number of points, the logarithmic divergence ha been
smoothened by the ultraviolet cutoff, but apart from that the two
functions are quite similar. The choice of different values of
$z'$ gives similar pictures.
\begin{figure}[bpht]
\epsfxsize=2.5 in \centerline{\epsfxsize=2.5
in\epsffile{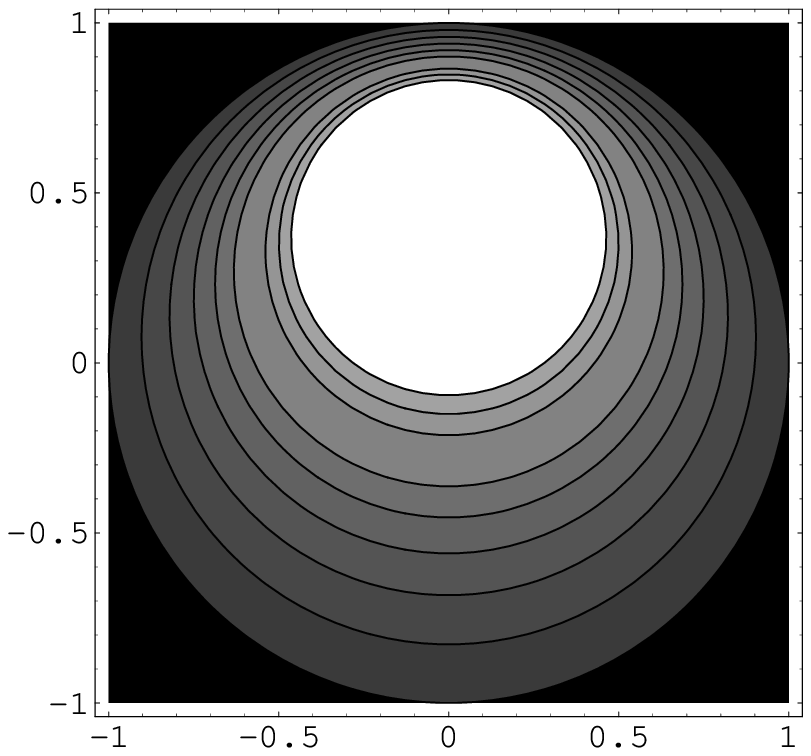} \epsfxsize=2.5
in\epsffile{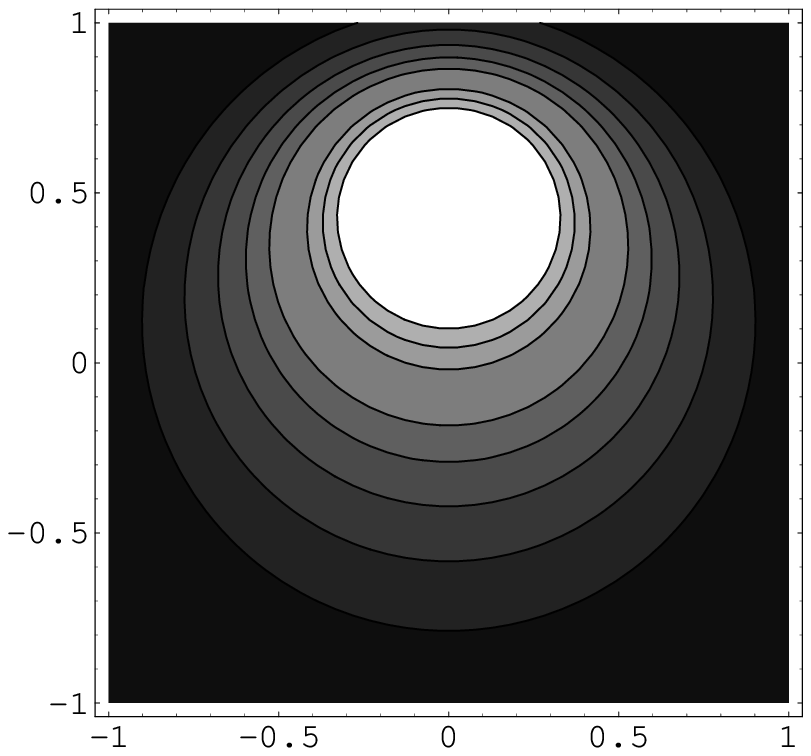}} \caption{\baselineskip=12pt {\it The
density plot of the Green's functions with the same parameters as
in figure~\ref{3d}. The exact function is on the left.  }}
\bigskip
\label{densityplot}
\end{figure}
\section{Conclusions and Outlook}
\setcounter{equation}{0}

We have seen as the fuzzy approximation to the disc works fairly
well for the calculations we have attempted, which were based on
the structure of the algebra and on the Laplacian. The first
conclusion that can certainly be drawn is that the fuzzy disc, as
a drum, sounds very much like a commutative disc even with an
approximation based on relatively small matrices. We performed
only one exact analytical calculation as a comparison with the
continuum case, but the method can be extended to more involved
calculations where it can be a very useful tool. Although the
approximation is finite the resulting (fuzzyfied) functions,
correlation functions etc.\ are defined for all points on the
disc, and rotational invariance is maintained, thus enabling
better comparisons with the continuum theory.

Even though the main stress of this paper has been on the
numerical approximations, the fuzzy disc is also an interesting
object from the mathematical point of view, and its setting on a
more rigorous footing would not only have an intrinsic value, but
also help understanding the role of the limits in the
approximation of field theory and renormalization.

\setcounter{section}{0}
\renewcommand{\thesection}{\Alph{section}}
\setcounter{equation}{0}

\section{\label{appcon}Conversions between different basis}
In this section we collect some useful conversions formulas
between the various basis. They are all standard expressions, but
care has to be taken because of the nonstandard commutation
relation~\eqn{aadcomm}. The resolutions of the identity are
\be
1=\sum_{n=0}^\infty \dm{n}{n}=\frac 1{\pi\theta} \int d^2z\;
\dm{z}{z} \ . \label{ident}
\ee
Coherent states are an overcomplete basis and they are not
orthogonal, their inner product is:
\be
\hs{z}{z'}=e^{-\frac{|z|^2+|z'|^2-2\bar zz'}{\theta}}  \ ,
\ee
while the inner products between a coherent state and an
eigenstate of $N$ are given by:
\bea
\hs{z}{n}&=&e^{-\frac{|z|^2}{2\theta}}\frac{\bar z
^n}{\sqrt{n!\theta^n}} \nn\\
\hs{n}{z}&=&e^{-\frac{|z|^2}{2\theta}}\frac{ z
^n}{\sqrt{n!\theta^n}} \ . \label{znsca}
\eea
Known $\pt_{mn}$ is possible to calculate $\varphi_{mn}$ from
\be
\varphi_{lk}=\sum_{q=0}^{\min(l,k)}\pt_{l-q\
k-q}\frac{\sqrt{l!k!\theta^{l+k}}}{\theta^qq!}   \ ,
\label{taylortodm}
\ee
while the converse is given by
\be
\pt_{mn}=\sum_{p=0}^{\min(n,m)} \frac{(-1)^p\varphi_{m-p\
n-p}}{p!\sqrt{(m-p)!(n-p)!\theta^{m+n}}}
 \label{dmtotaylor}
\ee

\subsection*{Acknowledgments}
We thank A.P.~Balachandran, B.~Dolan, G.~Immirzi, G.~Landi,
G.~Marmo, X.~Martin, D.~O'Connor, P.~Presnajder, R.~Szabo,
A.~Stern and J.~Varilly for useful discussions. In particular
D.~O'Connor's questions on the spectrum of the Laplacian and
G.~Landi's on the nature of the subalgebra helped very much the
sharpening of our ideas. We especially thank A.P.~Balachandran,
K,~Gupta and S.~K\"urk\c{c}\"{u}o\v{g}lu for communicating to us
their work before publication, and for sharing with us their
thoughts. This work has been supported in part by the {\sl
Progetto di Ricerca di Interesse Nazionale {\em SInteSi}}.

\end{document}